\documentclass[twocolumn, pra, showpacs,superscriptaddress]{revtex4}
\usepackage{amsfonts}
\usepackage{amsmath}
\usepackage{amssymb}
\usepackage{graphicx}

\begin{document}

\author{Jie Zhang}
\email{zhangjie01@tyut.edu.cn}
\affiliation{College of Physics and optoelectronics, Taiyuan University of Technology,
Taiyuan 030024, P. R. China}
\affiliation{Key Lab of Advanced Transducers and Intelligent Control System of Ministry
of Education, Taiyuan University of Technology, Taiyuan 030024, P. R. China}
\author{Xue Hou}
\affiliation{College of Physics and optoelectronics, Taiyuan University of Technology,
Taiyuan 030024, P. R. China}
\author{Bin Chen}
\affiliation{College of Physics and optoelectronics, Taiyuan University of Technology,
Taiyuan 030024, P. R. China}
\author{Yunbo Zhang}
\affiliation{Institute of Theoretical Physics, Shanxi University,
Taiyuan 030006, P. R. China}
\title{Fragmentation of a spin-1 mixture in a magnetic field}

\begin{abstract}
We study the ground state quantum fragmentation in a mixture of
a polar condensate and a ferromagnetic condensate
when subject to an external magnetic field.
We pay more attentions to the polar
condensate and find that it will be less fragile in the mixture when
perturbed by the magnetic field. Both atom numbers and the number fluctuations
in the spin-0 component will keep in a high magnitude of order of $N$ when the
magnetization of the system is increased. The role of the ferromagnetic
condensate is to provide a uniform and stable background which can delay the
rapid shrink of the 0-component population and make it possible to capture
the \textquotedblleft super-fragmentation \textquotedblright. Our
method has potential applications in measuring the inter-species spin-coupling interaction
through adjusting the magnetic field.
\end{abstract}

\pacs{03.75.Mn, 67.60.Bc, 67.85.Fg}
\maketitle

\section{Introduction}

Recent experimental breakthroughs in spinor Boson-Einstein Condensate, such
as the sub-Poissonian spin correlations generated by atomic four-wave spin
mixing \cite{Chapman.1}, the atomic squeezed states realized in the spin-1
ultracold atomic ensembles \cite{Chapman.2}, and the antiferromagnetic spatial
ordering observed in a quenched one-dimensional spin-1 gas \cite{raman}, are
all in connection with the vacuum fluctuations and recall attentions to
the finite particle number effect beyond the mean-field treatment. The
vacuum fluctuations become a significant subject in more and more
experimental facts, e.g., atomic quantum matter-wave optics, atomic spin
squeezing and quantum information. As one of the active frontiers, the
spin-1 ultracold atomic ensemble is often adopted. With the basic interaction form $V(%
\mathbf{r})=(\alpha +\beta \mathbf{F}\cdot \mathbf{F})\delta (\mathbf{r})$,
the properties of such a three-component spinor condensate \cite{HoOhmi} have
been demonstrated experimentally \cite{Stenger} and two different phases
reflecting fundamental properties of spin correlation are identified: the
so-called polar and ferromagnetic states for $\beta >0$ ($^{23}$Na) and $%
\beta <0$ ($^{87}$Rb) atomic condensates respectively. The mixture
of two spinor condensate with the ferromagnetic and polar atoms, respectively,
show more attractive quantum effects \cite%
{Luo,Xu09,Xu2,zhang1,zhang2, YuShi, WangDajun,WangDajun2}.
With the help of sympathetic cooling, the BEC mixtures of Na and Rb have been realized
and it is interesting to observe the interspecies interaction induced immiscibility between
the two condensates \cite{WangDajun}.

The ground state of the condensate with $\beta >0$ has been predicted to be either
polar ($n_{0}=N$) or antiferromagnetic ($n_{1}=n_{-1}=N/2)$ within
the mean-field treatment, where the condensate is usually described by a
coherent state. However, the many-body theory by Law, Pu and Bigelow \cite{Law}
pointed out that the ground state of $\beta >0$ atoms is a spin
singlet with properties ($n_{1}=n_{0}=n_{-1}=N/3$) drastically
different with the results predicted by the mean field theory. Soon, Ho
and Yip \cite{HoYip} show that this spin singlet state is a fragmented
condensate with anomalously large number fluctuations and thus has fragile
stability. The remarkable nature of this super-fragmentation is that the
single particle reduced density matrix gives three macroscopic eigenvalues
($N/3$) with large number fluctuations $\Delta n_{1,0,-1}\sim N$. Similar
considerations were also addressed by Koashi and Ueda \cite{Masato
Koashi,M.Ueda,M.Ueda2}. The signature of fragmentation is then refer to the
anomalously large fluctuations of the populations in the Zeeman levels. This
is a super-poissonian correlation character, and the large number
fluctuations shrink rapidly as the experimentally adventitious perturbations
exist, such as magnetic field or field gradient.

In this paper we will report the influence of external magnetic field on the
spinor condensate with $\beta >0$, but on the premise of doping many
ferromagnetic atoms in it. The interspecies spin coupling interaction
arises and we propose a valid procedure to observe and control the
fragmented states. If the ferromagnetic atoms in the mixture are condensed,
the ground state favors all atoms aligned along the same direction and
provides a uniform and stable background which can delay the rapidly
shrinking of the number fluctuations when the inter-species coupling
interaction is adjusted. The back action from polar atoms on to the more
stable ferromagnetic atoms is negligible. Doping ferromagnetic atoms into
spin-1 polar condensate can effectively influence the vacuum fluctuations
and will have potential applications in quantum information and
quantum-enhanced magnetometry.

\section{Hamiltonian of the mixture}

We consider the mixture of two spinor condensates of $N_1$ ferromagnetic and
$N_2$ polar atoms, respectively. The intra-condensate atomic spin-1
interaction takes the standard interaction form $V_{k}(\mathbf{r})=(\alpha
_{k}+\beta _{k}\mathbf{F}_{k}\cdot \mathbf{F}_{k})\delta (\mathbf{r})$
with $k=1,2$. The inter-condensate interaction between the ferromagnetic and
polar atoms is 
$V_{12}(\mathbf{r})=\frac{1}{2}(\alpha +\beta \mathbf{F}_{1}\cdot \mathbf{F}%
_{2}+\gamma P_{0})\delta (\mathbf{r})$, 
which is more complicated because collision can occur in the total spin $F_{%
\mathrm{tot}}=1$ channel between different atoms \cite{Luo,Xu09}. The
parameters $\alpha ,\beta $, and $\gamma $ are related to the $s$-wave
scattering lengths in the three total spin channels and the reduced mass $%
\mu $ for atoms in different species, and $P_{0}$ projects an inter-species
pair into spin singlet state. 
Within the single spatial-mode approximation (SMA) \cite{Law,Pu,Yi} for each
of the two spinor condensates, the spin-dependent Hamiltonian for the
mixture finally reads as
\begin{equation}
\hat{H}=c_{1}\beta _{1}\mathbf{\hat{F}}_{1}^{2}+c_{2}\beta _{2}\mathbf{\hat{F%
}}_{2}^{2}+c_{12}\beta \mathbf{\hat{F}}_{1}\cdot \mathbf{\hat{F}}_{2}+\frac{%
c_{12}\gamma }{3}\hat{\Theta}_{12}^{\dag }\hat{\Theta}_{12},  \label{Ham}
\end{equation}%
where $\mathbf{\hat{F}}_{1}=\hat{a}_{i}^{\dag }\mathbf{F}_{1ij}\hat{a}_{j}$ (%
$\mathbf{\hat{F}}_{2}=\hat{b}_{i}^{\dag }\mathbf{F}_{2ij}\hat{b}_{j}$) are
defined in terms of the $3\times 3$ spin-1 matrices with $i(j)=1,0,-1$, and $%
\hat{a}_{i}^{\dag }(\hat{b}_{i}^{\dag })$ creates a ferromagnetic (polar)
atom in the hyperfine state $i$. The operator
\begin{equation}
\hat{\Theta}_{12}^{\dag }=\hat{a}_{0}^{\dag }\hat{b}_{0}^{\dag }-\hat{a}%
_{1}^{\dag }\hat{b}_{-1}^{\dag }-\hat{a}_{-1}^{\dag }\hat{b}_{1}^{\dag },
\end{equation}%
creates a singlet pair with one atom each from the two species, similar to
\begin{equation}
\hat{B}^{\dag }=(\hat{b}_{0}^{\dag })^{2}-2\hat{b}_{1}^{\dag }\hat{b}%
_{-1}^{\dag },
\end{equation}%
for intra-species spin-singlet pair \cite{HoYip,Masato Koashi} when $\beta
_{2}>0$. The interaction parameters are $c_{1}=\frac{1}{2}\int d\mathbf{r}%
\left\vert \Psi (r)\right\vert ^{4},$ $c_{2}=\frac{1}{2}\int d\mathbf{r}%
\left\vert \Phi (r)\right\vert ^{4}$ and $c_{12}=\int d\mathbf{r}\left\vert
\Psi (r)\right\vert ^{2}\left\vert \Phi (r)\right\vert ^{2},$ which can be
tuned through the control of the frequency of the trapping potential \cite%
{Xu09}.

\section{Fragmentation in the magnetic field}

\begin{figure}[tbp]
\includegraphics[width=3.0in]{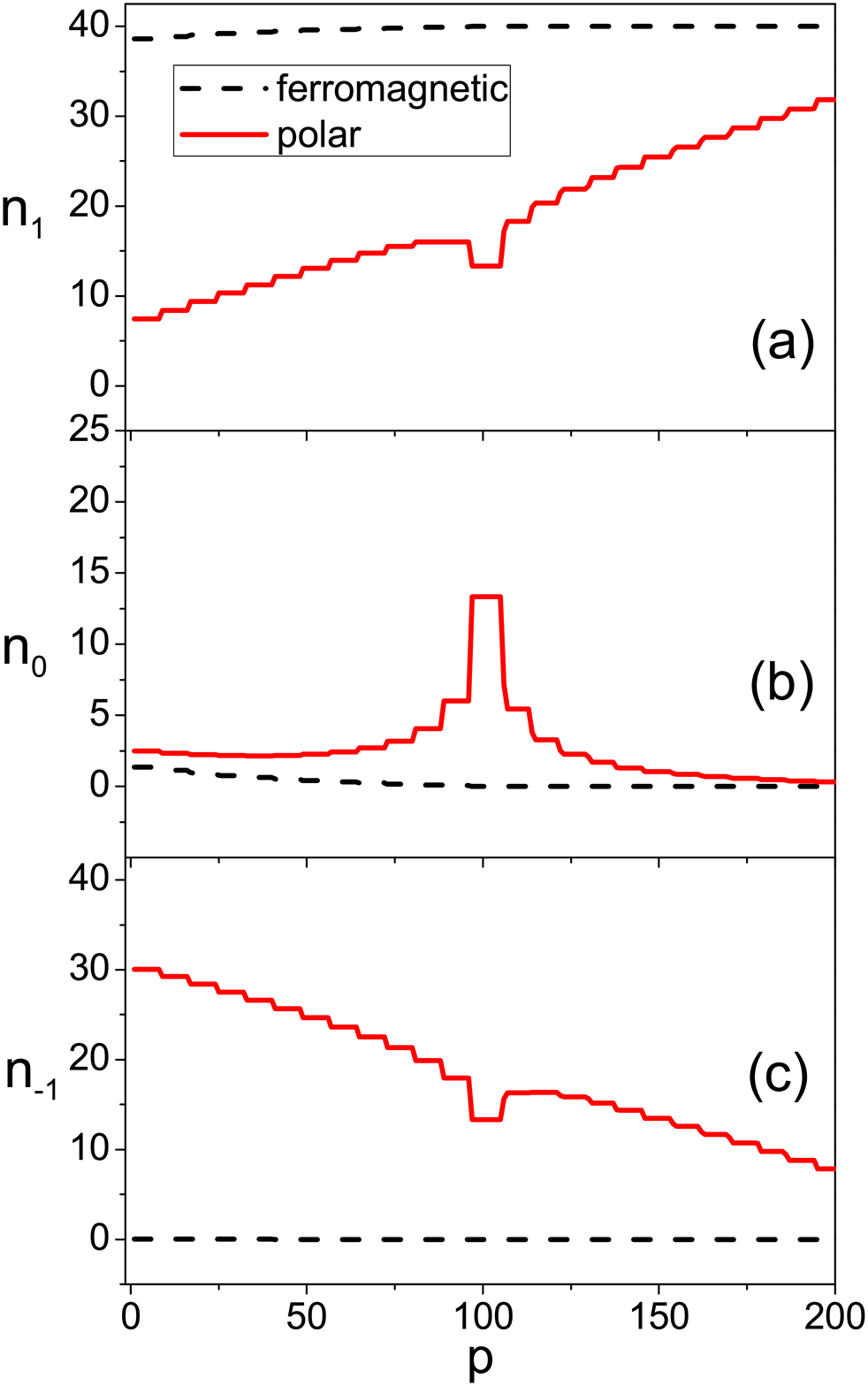}
\caption{(Color online) The dependence of atom numbers on $p$, at fixed
values of $c_{1}\protect\beta _{1}=-1$, $c_{2}\protect\beta _{2}=2$, and $%
c_{12}\protect\beta =2.5$. The total numbers of the two species are $N_{1}=
N_{2}=40$, and we consider the full-space with total magnetization $m$ a
variable. Black dashed and red solid lines denote the number distributions
in the ferromagnetic and polar condensate respectively. All interaction
parameters are in the units of $|c_{1}\protect\beta _{1}|.$}
\end{figure}

\subsection{Number distributions in a magnetic field}

When the interspecies scattering parameters are calculated in the degenerate
internal-state approximation (DIA) \cite{DIA0,DIA1,DIA2,DIA3}, the
low-energy atomic interactions can be mostly attributed to the ground-state
configurations of the two valence electrons, and the non-commutative term $%
\hat{\Theta}_{12}^{\dag }\hat{\Theta}_{12}$ can be neglected \cite{Luo,
Xu09, zhang1}. The ground states are classified into four distinct phases:
FF, MM$_{-}$, MM$_{+}$, and AA by three critical values of $c_{12}\beta={-}%
\frac{{(2N-1)}c_{2}\beta _{2}}{N}$, $0$, and $\frac{{(2N-1)}c_{2}\beta _{2}}{%
N+1}$ \cite{zhang1}.

In this paper we discuss the atom number distribution and fluctuation in an
external magnetic field. The spin-dependent Hamiltonian in the magnetic
field reads,
\begin{eqnarray}
\hat{H} &=&c_{1}\beta _{1}\mathbf{\hat{F}}_{1}^{2}+c_{2}\beta _{2}\mathbf{%
\hat{F}}_{2}^{2}+c_{12}\beta \mathbf{\hat{F}}_{1}\cdot \mathbf{\hat{F}}_{2}
\label{hhh} \\
&&-c_{1}p_{1}\hat{F}_{1z}-c_{2}p_{2}\hat{F}_{2z},  \notag
\end{eqnarray}%
where only the linear Zeeman terms are considered. As the SU(2) symmetry is
broken in a spinor mixture, one can not eliminate the linear Zeeman effect
through a spin rotation \cite{WXzhang}. Meanwhile the quadratic Zeeman
energy, typically 2 orders of magnitude weaker than the linear terms, is
negligible in the calculation of number distributions. Furthermore, we take $%
p=c_{1}p_{1}=c_{2}p_{2}$ in the following discussion, which can easily
realized through adjusting the trapping frequency.

We consider the direct product of the Fock states of the two species $%
\left\vert n_{1},n_{0},n_{-1}\right\rangle _{1}\otimes \left\vert
n_{1},n_{0},n_{-1}\right\rangle _{2}$, and do not restrict the model in the
subspace with zero total magnetization \cite{zhang1,zhang2}. Instead, we
consider the full space including all possible system magnetization $%
m=m_{1}+m_{2}$. Using the full quantum approach of exact diagonalization, we
study the response of the two species to the external magnetic field $p$ for
$N_{1}=N_{2}=40$. The three critical points for the phase boundaries are
approximately $c_{12}\beta=-4,0,4$.

The field dependence of the population is shown in Fig. 1 for the MM$_{+}$
phase at $c_{12}\beta =2.5$, where polar atoms are partly polarized in the
oppsite direction as the ferromagnetic atoms \cite{zhang1}. We notice that
the ferromagnetic atoms (black dashed lines) are very sensitive to the
magnetic field, i.e. atoms quickly redistribute in the $n_{1}$ component and
the magnetization of ferromagnetic condensate $m_{1}=n_{1}-n_{-1}$ saturates
immediately. The ferromagnetic atoms actually form a stable condensate and
provide a uniform magnetic background in the mixture. The polar atoms
present a stepwise increase (decrease) in the atom number distribution $%
n_{1}(n_{-1})$ when the field increases. For small $p$ and positive $%
c_{12}\beta $, the system favors a negative magnetization ($%
m_{2}=n_{1}-n_{-1}$) of polar condensate, and $m_{2}$ will reverse and tend
to saturate for large magnetic field. We notice that the super-fragmented
state featured with $n_{1}=n_{0}=n_{-1}=\frac{N}{3}$ remarkably arises
around a special value of $p=100$.

The situation becomes more simple if the parameter $c_{12}\beta$ is
negative, that is, in the FF phase (or MM$_{-}$ phase), where polar atoms
are fully (partly) polarized in the same direction as the ferromagnetic
atoms. The enhanced ferromagnetic effect and the external magnetic field
jointly suppress the atom number distribution $n_{0}$ and $n_{-1}$ of the
polar condensate to zero, and at the same time saturate $n_{1}$ and the
magnetization $m_{2}$ without magnetization reversal.

\begin{figure}[tbp]
\includegraphics[width=2.7in]{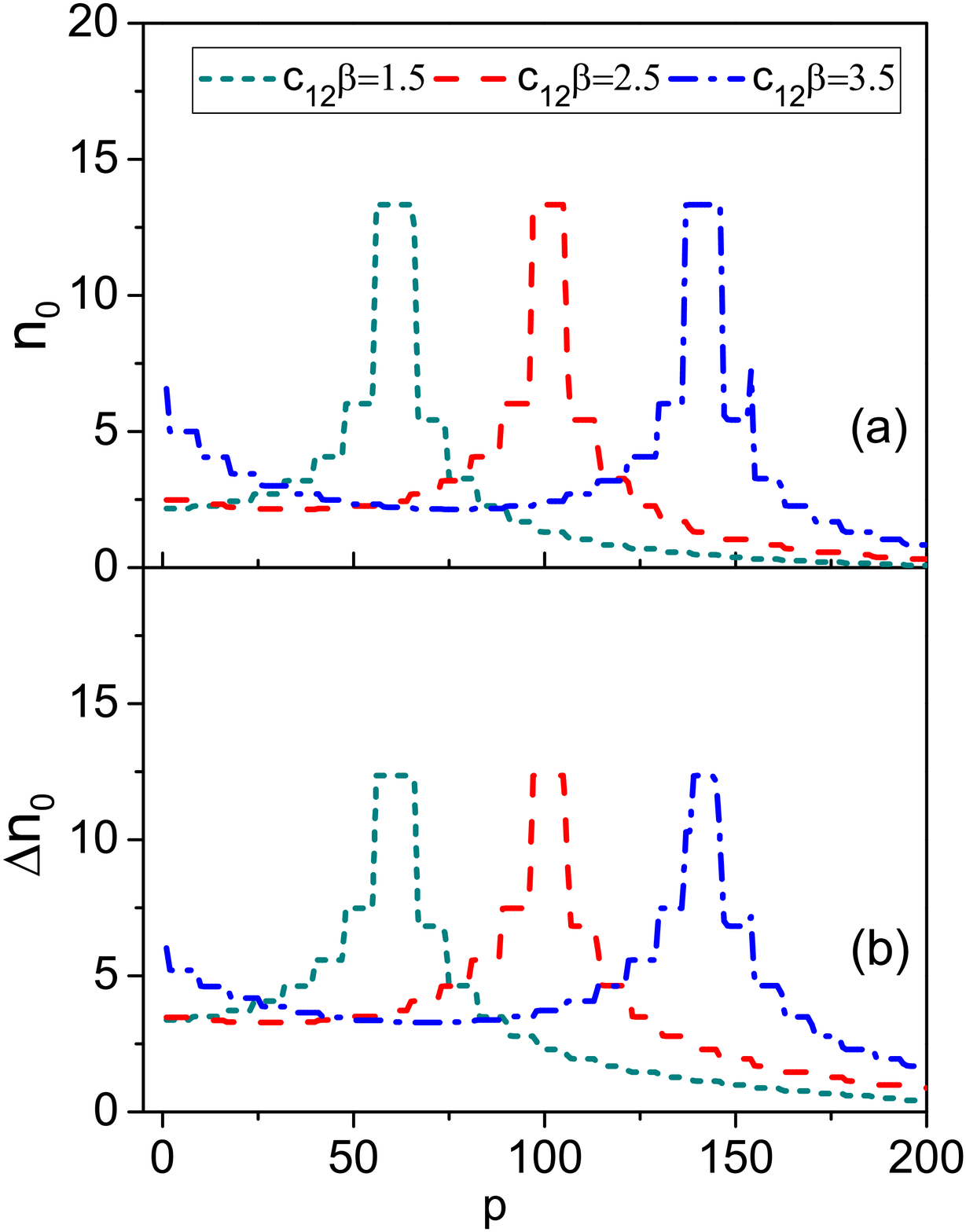}
\caption{(Color online) The dependence of atom numbers and fluctuations on $%
c_{12}\protect\beta $ and $p$ at fixed values of $c_{1}\protect\beta _{1}=-1$
and $c_{2}\protect\beta _{2}=2$. This column only shows the results of polar
atoms with $n_{0}$ and $\Delta n_{0}$. When the extra magnetic field
parameter $p$ (in the units of $|c_{1}\protect\beta _{1}|$) increases, there
are serval critical points associated with $c_{12}\protect\beta $. All
interaction parameters are in the units of $|c_{1}\protect\beta _{1}|.$}
\end{figure}

\begin{figure}[tbp]
\includegraphics[width=2.5in]{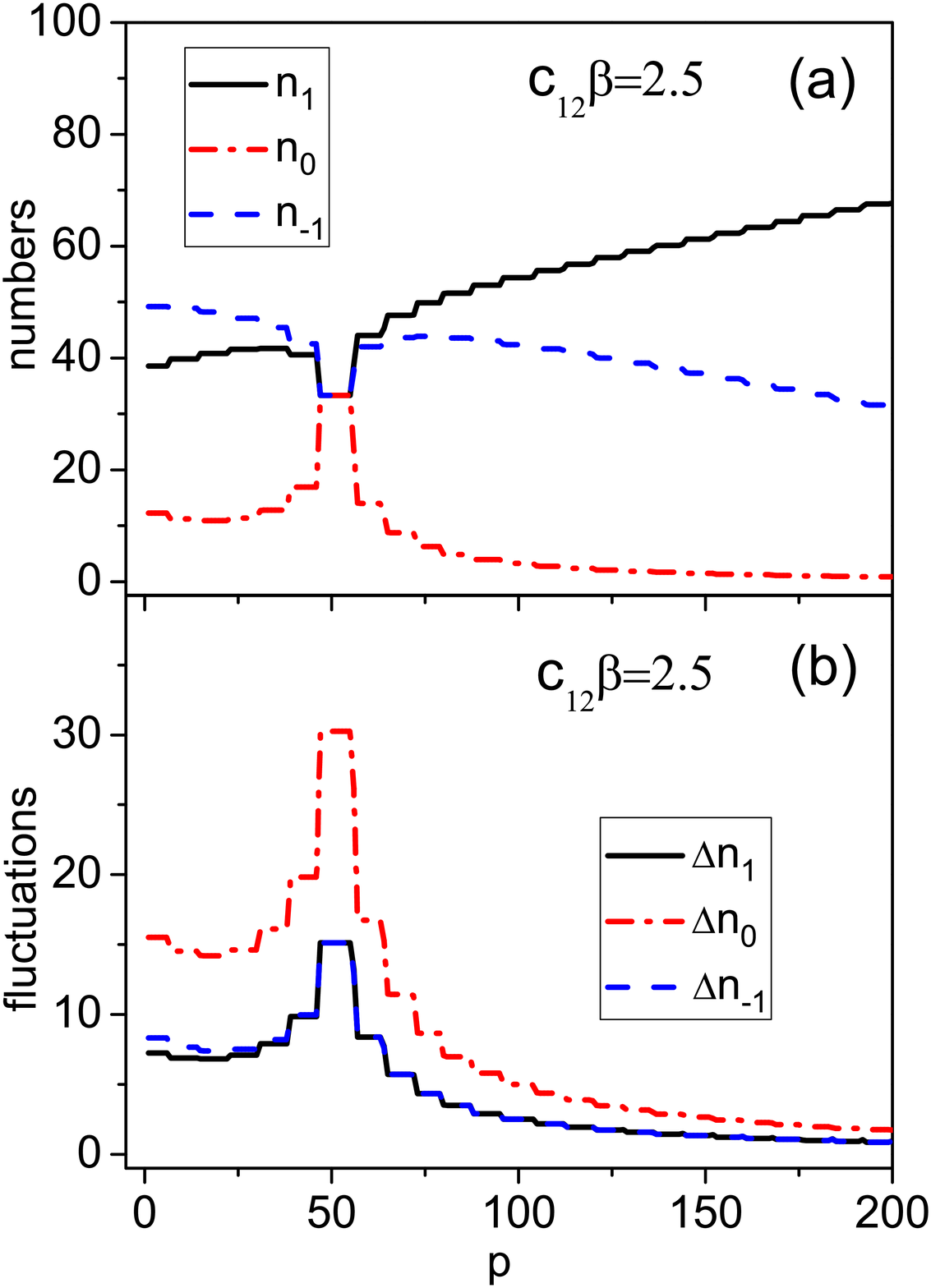}
\caption{(Color online) The dependence of atom number distributions and
number fluctuations in the polar condensate on magnetic coefficient $p$ at
fixed values of $c_{1}\protect\beta _{1}=-1$, $c_{12}\protect\beta =2.5$ and
$c_{2}\protect\beta _{2}=2$. The total numbers of the two species are $N_{1}
=20$, $N_{2}=100$. Black
solid, red dash-dot, and blue dashed lines denote atom numbers and the fluctuations on
the $1, 0,$ and $-1$ sub-levels respectively. All interaction parameters are in
the units of $|c_{1}\protect\beta _{1}|.$}
\end{figure}

\subsection{Retrieving the super-fragmented state}

The super-fragmented state, with equal population $n_{1,0,-1}=N/3 $ and zero
magnetization $m_{2}=0$, has been predicted in the pure spin-1 polar
condensate and described by a spin singlet state in the form $\propto (\hat{B%
}^{\dag })^{N/2}\left\vert 0\right\rangle $ \cite{HoYip}. $\hat{B}^{\dag }(%
\hat{B})$ is invariant under spin rotations and commutes with $\mathbf{\hat{F%
}}_{2}$\ and $\hat{F}_{2z}$. In the ground state of the system subject to an external
magnetic field $(\hat{b}_{1}^{\dag })^{m_{2}}(\hat{B}^{\dag })^{%
\frac{N_{2}-m_{2}}{2}}\left\vert 0\right\rangle $, one can see a rapid shrink
of the spin-0 component distribution $n_{0}$ and an unbalanced population $%
m_{2}>0 $. The super-fragmented state then reduces to a much more generic
fragmented state: a two component number state with essentially zero
fluctuations. Such state with polar interaction was not likely realized in
typical experiments due to its fragility towards any perturbation breaking
spin rotational symmetry.

For the spin-1 polar condensate doped with many ferromagnetic atoms,
we can retrieve this super-fragmented state in the presence of an external
field. For some special values of the magnetic field, both the spin-0 component
population and number fluctuations would not shrink but revive to macroscopic
orders of the super-fragmented state. In Fig. 2, we illustrate the revival points for
three inter-species coupling parameters $c_{12}\beta $. These revival points are
found to move towards larger value of $p$ as $c_{12}\beta $ increases.

As learned from previous studies \cite{Yi}, the mean-field treatment is
efficient for atomic interaction of the ferromagnetic type. The much more
stable ferromagnetic condensate in the mixture can be formulated in the
mean field treatment as a boson-enhanced effective magnetic field. This
simplifies the Hamiltonian (\ref{hhh}) as
\begin{eqnarray}
\hat{H} &=&c_{1}\beta _{1}\langle \mathbf{\hat{F}}_{1}^{2}\rangle
+c_{2}\beta _{2}\mathbf{\hat{F}}_{2}^{2}+c_{12}\beta \langle \mathbf{\hat{F}}%
_{1}\rangle \cdot \mathbf{\hat{F}}_{2} \\
&&-c_{1}p_{1}\langle \hat{F}_{1z}\rangle -c_{2}p_{2}\hat{F}_{2z}  \notag \\
&=&c_{2}\beta _{2}\mathbf{\hat{F}}_{2}^{2}+A\hat{F}_{2z}+C  \notag
\end{eqnarray}%
where $\langle \mathbf{\hat{F}}_{1}\rangle =\langle \hat{F}%
_{1z}\rangle =N_{1},$ $A=c_{12}\beta N_{1}-c_{2}p_{2},$ $C=c_{1}\beta
_{1}N_{1}(N_{1}+1)-c_{1}p_{1}N_{1}.$ The criterion for the emergence of
super-fragmented state is $p=c_{12}\beta N_{1}$, where the magnetic field ($p$),
the optical trapping frequency ($c_{12}$), and the number of the doped
ferromagnetic atoms ($N_{1}$) are all adjustable. When the magnetic
field matches the condition that $c_{12}\beta N_{1}$ and $c_{2}p_{2}$ cancel
each other, we may achieve the super-fragmented state
in a magnetic field. The three critical points in Fig. 2
are found to agree with the numerical results exactly.

Next, we turn to the situation with population imbalance in the two species.
Fig. 3 illustrates the location of the critical point when the inter-species
coupling parameter $c_{12}\beta $ is fixed to be 2.5 and the atom numbers
for the two species are $N_{1}=20$ and $N_{2}=100$. As the mean-field
picture works well for the ferromagnetic atoms, we still get the crucial point
$p=2.5\times N_1$ in Fig. 3. When equal population $n_{1}=n_{0}=n_{-1}=
N/3$ occurs for the polar condensate, the number fluctuations also
instantaneously reach to the macroscopic levels. Our numerical results for the
fluctuation relation ($\Delta n_{0}=2\Delta n_{\pm 1}$) agree exactly
with the algebraic results in \cite{HoYip} for pure polar condensate.
With the emergence of equal population $N/3$ regarded as a sign
of anti-ferromagnetic spin interaction, the inter-species spin coupling
interaction $c_{12}\beta$ can be estimated by the location of the critical magnetic
field.

\begin{figure}[tbp]
\includegraphics[width=2.5in]{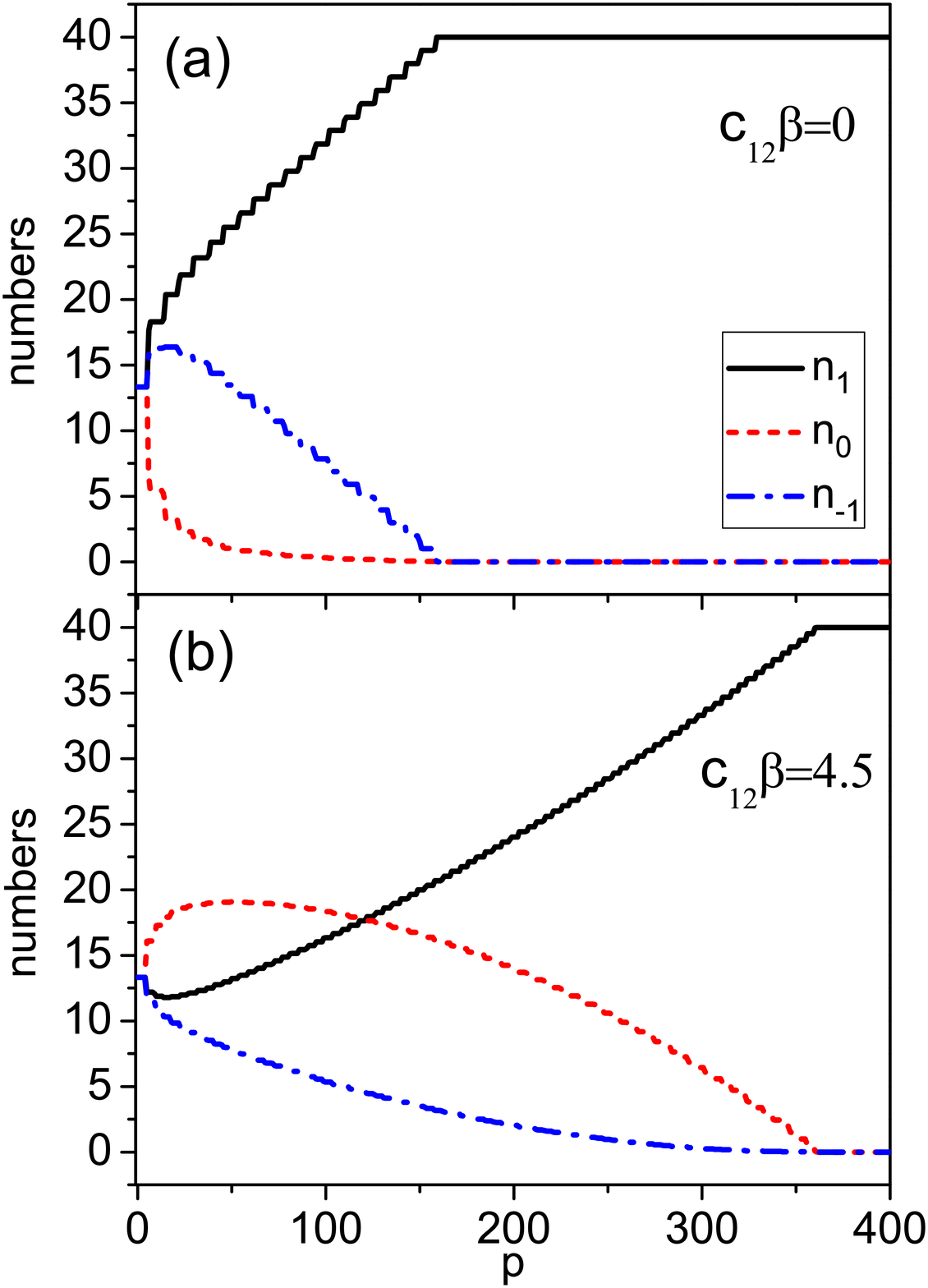}
\caption{(Color online) The dependence of atom number distributions $%
n_{1,0,-1}$ of the polar condensate on both $c_{12}\protect\beta $ and
magnetic field $p$ at fixed values of $c_{1}\protect\beta _{1}=-1$ and $%
c_{2}\protect\beta _{2}=2$. The total numbers of the two species are $N_{1}=%
N_{2}=40$. Black solid, red dot, and blue dash-dot lines denote the numbers
on the 1, 0, and $-1$ sub-levels respectively. All interaction parameters are
in the units of $|c_{1}\protect\beta _{1}|.$}
\end{figure}

\subsection{ AA phase in a magnetic field}

AA phase is another super-fragmented state which have been predicted
in the absence of magnetic fields \cite{zhang1}. It is a many-body spin singlet, which
requires exactly the same atoms number of the two species ($N=N_{1}=N_{2}$),
and total spins from different species polarized in opposite
directions. In the notation of the angular momentum representation
\begin{equation}
\left\vert F_{1},F_{2},F,m\right\rangle =\sum
C_{F_{1,}m_{1};F_{2,}m_{2}}^{F,m}\left\vert F_{1},m_{1}\right\rangle
\left\vert F_{2},m_{2}\right\rangle ,
\end{equation}
AA phase is denoted as $\left\vert \phi _{AA}\right\rangle=\left\vert N,N,0,0
\right\rangle$ with $F_1,F_2$ and $F$ the total spin quantum numbers of the ferromagnetic
atoms, polar atoms, and the mixture and $m_1,m_2$ and $m$ the corresponding
$z$-components. The intra-species angular momentum states involved in the AA
phase, $\left\vert N,m_{1}\right\rangle $ and $\left\vert
N,m_{2}\right\rangle $, should obey the constraint $m_{1}+m_{2}=0$.
The interesting feature of AA phase is the equal
distribution of atoms in the six components ($N/3$) and the large number fluctuations. Using
the full quantum approach of exact diagonalization, and considering the full
space including all possible system magnetization $m=m_{1}+m_{2}$, we study
the responses of the AA phase to the external magnetic field with $%
N_{1}=N_{2}=40$, and compare the results with the super-fragmented state in
the pure condensate \cite{HoYip}.

In Fig. 4, we compare these two typical fragmented ground states: the
inter-species entangled singlet $\left\vert \phi _{AA}\right\rangle $ and
the pure polar singlet $(\hat{B}^{\dag })^{\frac{N}{2}}\left\vert
0\right\rangle $, which belong to two special phases characterized by
typical values of the interaction parameter: $c_{12}\beta=4.5$, and $%
c_{12}\beta=0$. We find that the responses of the $n_{0}$ component to the
magnetic field are quite different. Fig. 4a shows the influence of a
magnetic field on a pure polar condensate. As $p$ increases, the 0-component
distribution $n_{0}$ (red dashed line) shrink rapidly, which agree with the
algebraic results in \cite{HoYip} $n_{0}=(N_{2}-m_{2})/(
2m_{2}+3)$. Fig. 4b shows the influence of a
magnetic field on the polar atoms in the AA
phase. We find that $n_{0}$ does not shrink rapidly in the beginning,
instead, it increases first and remains in a high value for a certain range of $p$.
Compared with the pure polar spin-singlet state, the inter-species entangled
singlet seems to be more difficult to be magnetized as $p$ increases.
When all the polar atoms are forced to arrange in the opposite
direction of the ferromagnetic ones, the bose-enhanced magnetic background
formed by ferromagnetic condensate will be canceled. The system
spontaneously breaks into a high symmetry state when the interaction
parameter $c_{12}\beta$ goes across the phase transition point $\frac{{%
(2N-1)}c_{2}\beta _{2}}{N+1}$. The magnetic behaviors of the two species are
then identical (except for the atom mass), and both ferromagnetic atoms and
polar atoms are equally to be magnetized when the field intensity is
increased. Compared with the pure singlet in Fig. 4a, the involved total
atoms are doubled, and at least a field strength $p\sim 400$ is needed to saturate the
magnetization. The external magnetic field can be used to characterize these two
spin-singlets through tracing the atoms numbers of $n_{0}$ component of the polar atoms.

\subsection{The inter-species $P_{0}$ effect}

If we refer to more general case beyond the DIA approximation, the $\gamma $
term of the Hamiltonian (\ref{Ham}) should be considered. We notice that%
\begin{equation}
\left[ \mathbf{\hat{F}}_{1,2}^{2},\hat{\Theta}_{12}^{\dag }\hat{\Theta}_{12}%
\right] \neq 0,\left[ \mathbf{\hat{F}}^{2},\hat{\Theta}_{12}^{\dag }\hat{%
\Theta}_{12}\right] =0,
\end{equation}%
which means in general they do not belong to a set of commutative operators.
However, we can numerically study the phase transition through the order
parameter $\langle \hat{\Theta}_{12}^{\dag }\hat{\Theta}_{12}\rangle$ \cite%
{zhang2}. To see more clearly the role played by the parameter $c_{12}\gamma$
on the fragmentation, we numerically diagonalize the Hamiltonian (\ref{Ham})
with $N_{1}=N_{2}=40$.

\begin{figure}[tbp]
\includegraphics[width=2.7in]{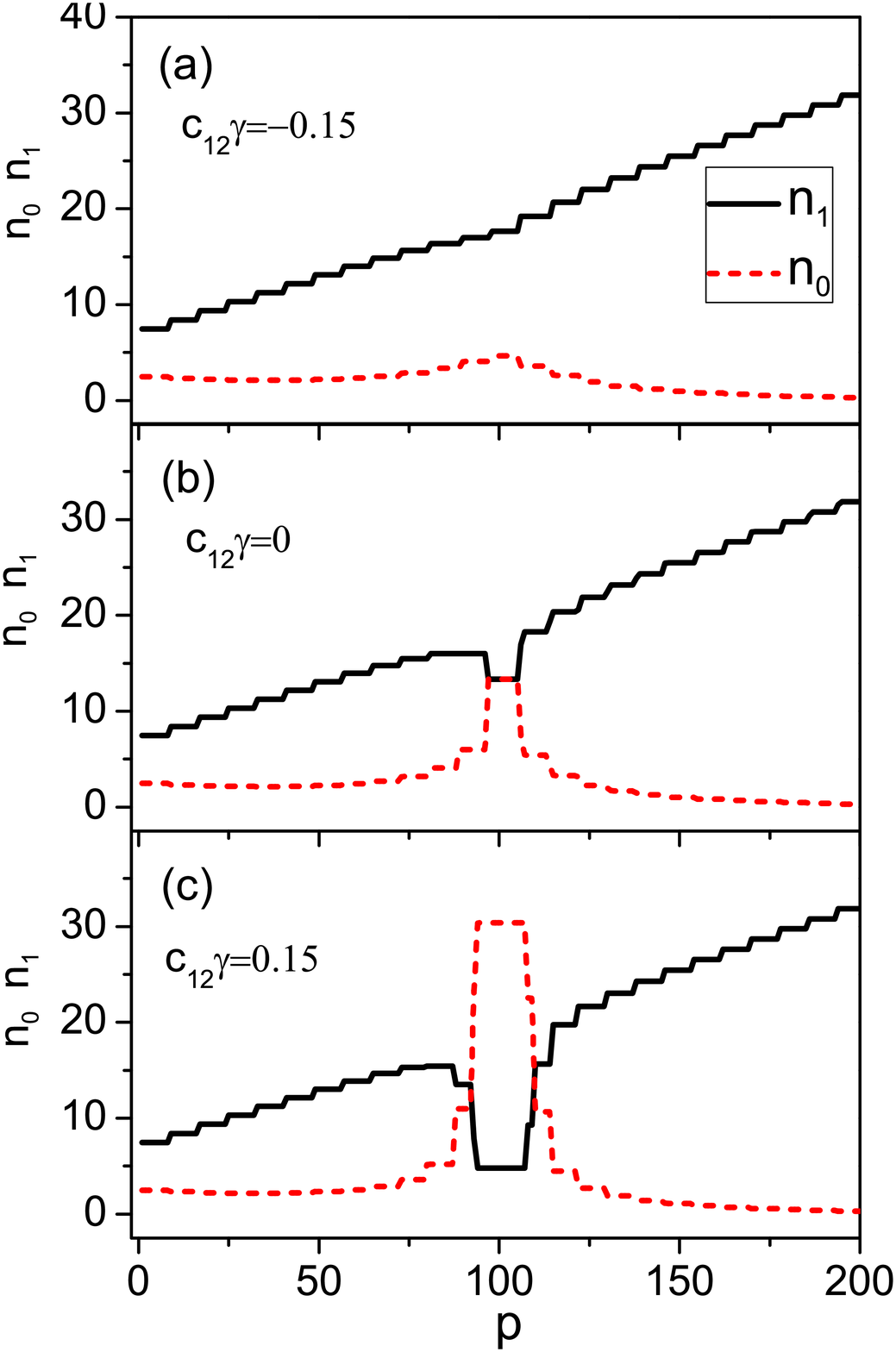}
\caption{(Color online) The dependence of atom number distribution $n_{1}$
and $n_{0}$ in the polar condensate on magnetic coefficient $p$ and $c_{12}%
\protect\gamma $ at fixed values of $c_{1}\protect\beta _{1}=-1$, $c_{12}%
\protect\beta =2.5$ and $c_{2}\protect\beta _{2}=2$. The total numbers of
the two species are $N_{1}=N_{2}=40$. Black solid line and red dashed
line denote the value of $n_{1}$ and $n_{0}$ respectively. All interaction
parameters are in the units of $|c_{1}\protect\beta _{1}|.$}
\end{figure}
In Fig.5, we illustrate the influence of a small $c_{12}\gamma \neq 0$ to
the population $n_{0} $ and $n_{1}$ of super-fragmented state which has been retrieved
in MM$_+$ phase in the presence of the magnetic field. We find that the
crucial point is still located at $p=2.5\times N_1$, but a tiny $c_{12}\gamma =0.15$
will elevate the $n_{0}$ component to a dominated value, meanwhile suppress
the $n_{1}$ and $n_{-1}$ components to lower level. The high occupation on $%
n_{0}$ component is an evidence of the nematic order \cite{Chapman.2}, but
the signature of fragmentation is still obvious. Away from for the critical
point where $n_{\pm1}$ are suppressed to be nearly zero, the $n_{1}$ ($n_{-1}$)
component is linearly increased (decreased) to meet the conditions that the
system magnetization should be increased. For $c_{12}\gamma =-0.15$, both $%
n_{0}$ and $\Delta n_{0}$ shrink, the system reduced to a much more generic
fragmented state with the system magnetization $m_{2}=n_{1}-n_{-1}$
which increases linearly with $p$. This situation is more like the pure polar condensate in
a weak magnetic field.
\begin{figure}[tbp]
\includegraphics[width=2.6in]{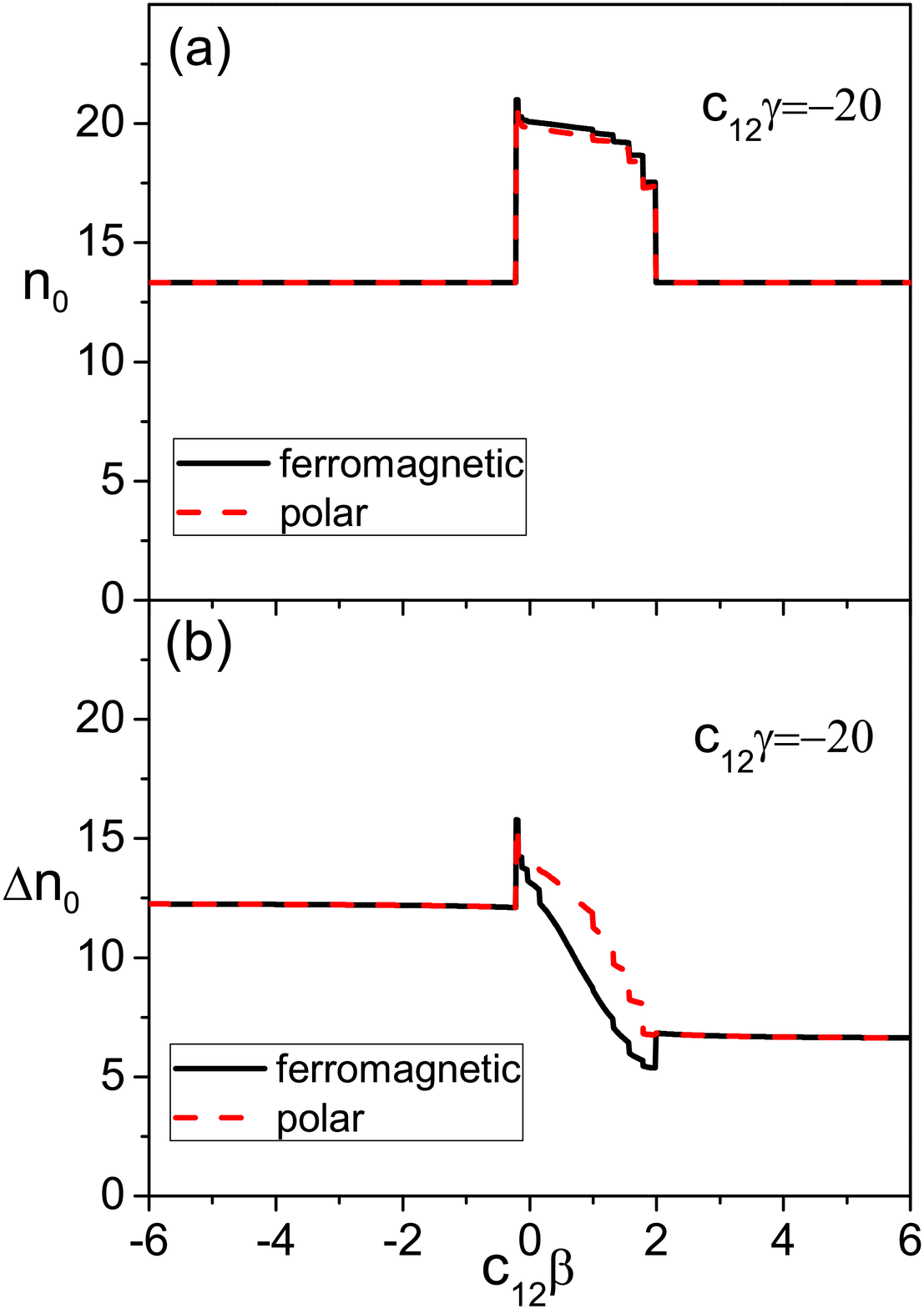}
\caption{(Color online) The dependence of atom numbers and fluctuations on $%
c_{12}\protect\beta $, $c_{12}\protect\gamma $ at fixed values of $p=0$, $%
c_{1}\protect\beta _{1}=-1$ and $c_{2}\protect\beta _{2}=2$. This graph only
shows the results of $n_{0}$ and $\Delta n_{0},$ when the interaction
parameter $c_{12}\protect\gamma $ equals to $-20$. The total numbers of the
two species are $N_{1}=N_{2}=40$ and we restrict the problem in full-space
without the external magnetic field. Black solid lines and red dash lines denote
the ferromagnetic and polar condensate respectively. All interaction
parameters are in the units of $|c_{1}\protect\beta _{1}|.$}
\end{figure}

The negative $\gamma $ term encourages pairing two
different types of atoms into singlets \cite{zhang2}. In
Fig. 6, the influences of $negative$ singlet pairing coefficient $%
c_{12}\gamma $ on the numbers and quantum fluctuations of the two species
are illustrated. We notice that the typical $N/3$ number
distributions arise both in the $c_{12}\beta <0$ and $c_{12}\beta >0$
regions when $c_{12}\gamma $ reaches to $-20$. The number fluctuation $%
\bigtriangleup n_{0}$ gives two steady values, which represent two typical
fragmented ground state: the inter-species entangled fragmented state for $%
c_{12}\beta >2$, and the inter-species independent fragmented state for $%
c_{12}\beta <0$. The fluctuations for these two states
\begin{eqnarray*}
\Delta n_{0} &=&\frac{\sqrt{N^{2}+9N}}{3\sqrt{5}},c_{12}\beta >2, \\
\Delta n_{0} &=&\frac{2\sqrt{N^{2}+2N}}{3\sqrt{5}},c_{12}\beta <0,
\end{eqnarray*}%
are found to match the numerical results in Fig. 6.

\section{Conclusion}

To conclude, we studied the ground state properties of a binary mixture of
ferromagnetic and polar spinor condensates in a magnetic field. Using the
full quantum approach of exact diagonalization, we can study the competition
between magnetic linear Zeeman effect and the inter-species spin coupling
interaction $c_{12}\beta$. The large vacuum fluctuation of number
distributions on the three zeeman levels inside the polar condensate is
worthy of investigation. We point out that the fragmentation properties of
polar condensate can be adjusted through the magnetic field (p), trapping
frequency $(c_{12})$, and the number of doped ferromagnetic atoms ($N_{1}$).
The ferromagnetic condensate is involved to provide a uniform and stable
background which can delay the rapidly shrinking of the large number
fluctuations. We illustrated the influences of the magnetic parameter $p$,
and identified two typical fragmented state with total spin $\langle
\mathbf{\hat{F}}^{2}\rangle =0$. The
positive inter-species spin coupling interaction ($c_{12}\beta>0 $) can
effectively entangle the different species, while for $c_{12}\beta<0 $ the
different species on their $F=1$ manifold are essentially independent. We
propose a possible mechanism to effectively measure the
inter-species spin coupling interactions through applying a magnetic field, as well as
discriminate the two types of many-body spin singlets.
Our work highlights the significant promises for experimental work on sodium
and rubidium atomic condensate mixtures and provide some useful information
for the study of photo-association of heteronuclear molecules.

This work is supported by the NSF of China under Grant Nos. 11204204, 11347181, the NSF of Shanxi Province under Grant Nos. 2012021003-2, 2014021011-1
and the fund of Taiyuan University of Technology for young teachers. YZ is also supported by
the National Basic Research Program of China (973 Program) under Grant No. 2011CB921601,
Program for Changjiang Scholars and Innovative Research Team in University (PCSIRT)(No. IRT13076).

\end{document}